\documentclass[12pt]{revtex4}
\usepackage{amsmath}
\usepackage{amsfonts}
\usepackage{amssymb}
\usepackage{graphicx}
\begin{document}

\title{Nucleus Driven Electronic Pulsation}
\author{H. Ludwig $^{a,b,c}$, R. Ruffini $^{a,b,c}$, S.-S. Xue $^{a,c}$\\
\small\itshape $^a$ICRANet, Pescara, Italy\ \ \ \ \ \ $^b$ICRANet, Nice, France\\
\small\itshape $^c$Dipartimento di Fisica and ICRA, Sapienza Universit\`a di Roma, Rome, Italy}
\begin{abstract}{We derive and solve by the spectral method the equations for a neutral system of ultra-relativistic electrons that are compressed to the radius of the nucleus and subject to a driving force.
This driving force can be thought of as originating from a nuclear breathing mode, a possibility we discuss in detail.}\end{abstract}
\maketitle

\section{Introduction}

The creation of electron-positron pairs through strong fields, which as been predicted theoretically by F. Sauter as early as 1931 \cite{sauter} but not yet been observed,
is nowadays of high interest for both laboratory scientists and astrophysicists \cite{preparata,ipp}. The obstacle to be solved is the creation of a sufficiently strong electric field.
In this note we will present an approach to investigate the dynamic response of an ultra-relativistic compressed electron gas subjected to the breathing mode of the underlying nucleus,
a system that promises to generate strong electric fields, depending on the mutual dynamics of negative and positive charges.
This model can, with slight modification like an adjustment of proton density, be extrapolated to astrophysical scales. In \cite{ludwig} we presented the stationary solutions
of pulsating electron gas in the framework of the Thomas-Fermi model, and we will now recapitulate the main results that are of importance for the present work. In this model the nucleus
is modeled as a sphere of homogenous positive charge distribution. The electrons are considered a charged perfect fluid with the equation of state of a Fermi gas at zero temperature.

\section{Spectral Method Formalism}

\begin{figure}[htp]
 \centering
\includegraphics[scale=0.9]{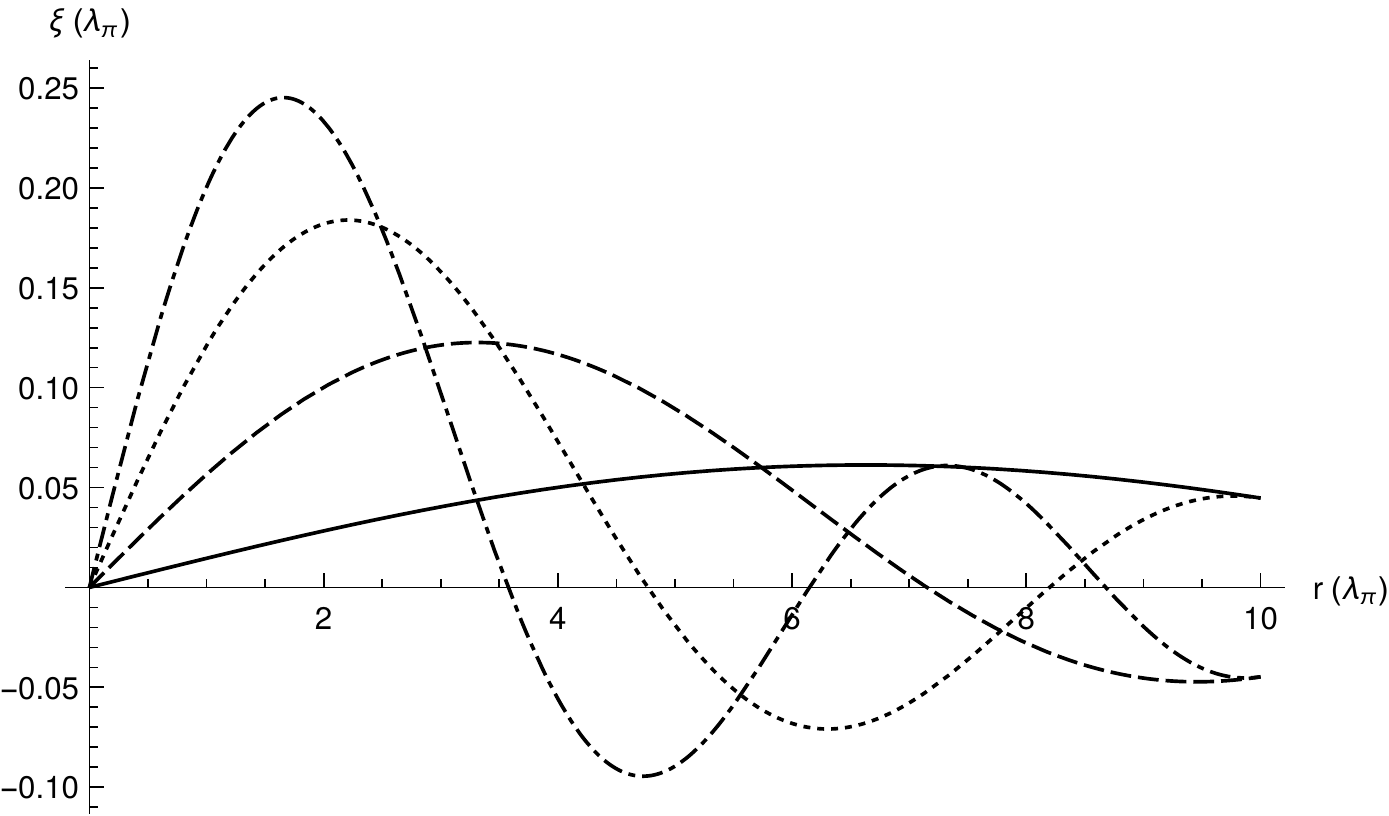}
 \caption{Basis functions}
 \label{fig:basisPlot}
\end{figure}

Since we want to treat the regime around nuclear density we can assume the ultra-relativistic limit
\begin{align}
 \rho=\frac{3}{4}(3\pi^2 n^4)^{1/3},\ \ \ \ \ p=\frac{1}{4}(3\pi^2 n^4)^{1/3}
\end{align}
for the equation of state, where $\rho$, $p$, and $n$ are energy density, pressure, and number density respectively of the electron gas. Throughout the text, natural units
$c=1$, $\hbar=1$ are used. When we assume the nucleus of proton number $Z$ to have a radius $r_\mathrm{nuc}=\Delta Z^{1/3}$ in units of the pion mass, and provide sufficient
external pressure to restrain the electrons to the volume of the nucleus, we obtain the very simple equilibrium configuration
\begin{align}
 \rho=\frac{9}{16\Delta^4}\left(\frac{3}{2\pi}\right)^{2/3},\ \ \ \ p=\frac{3}{16\Delta^4}\left(\frac{3}{2\pi}\right)^{2/3},\ \ \ \ n=n_\mathrm{nuc}=\frac{3}{4\pi\Delta^3}.
\end{align}
The parameter $\Delta$ measures the departure of proton density from nuclear densities. Positive and negative charges completely cancel, so there is no electric field in equilibrium.
Introducing a time-dependent displacement field $\xi(t,r)$ and treating Euler's and Maxwell's equations to first order, one obtains for stationary modes the second order
ordinary differential equation 
\begin{align}\label{diffEqn}
 \xi''(r)+\frac{2}{r}\xi'(r)+\left[3\omega^2-\left(\frac{12}{\pi}\right)^{1/3}\frac{3\alpha}{\Delta^2}-\frac{2}{r^2}\right]\xi(r)=0.
\end{align}
Here $\alpha$ is the fine structure constant, and $\omega$ is the frequency of the mode that has to be choosen according to approriate boundary conditions. This leads to
\begin{align}\label{omegaN}
 \omega_n=\left[\left(\frac{12}{\pi}\right)^{1/3}\!\!\frac{\alpha}{\Delta^2}+\frac{\pi^2 n^2}{3\,r_\mathrm{nuc}^2}\right]^{1/2}
  \ \ \ \mathrm{and}\ \ \ \ \xi_n(x)={\mathcal C}\,j_{1}\left(\pi n\,r/r_\mathrm{nuc}\right),
\end{align}
where $\mathcal C$ is an arbitrary factor and $j_1(z)$ is the first member of the spherical Bessel functions of first kind.
The first member of the spherical Bessel functions of first kind $j_1(z)$ allows to construct an orthonormal basis (see Figure \ref{fig:basisPlot})
\begin{align}
 \xi_n(r)=n\,\pi\,2^\frac{1}{2}\,r_\mathrm{nuc}^{-\frac{3}{2}}\,j_1(n\,\pi\,r/r_\mathrm{nuc}),\ \ n=1,2,3...
\end{align}
on the interval $[0,r_\mathrm{nuc}]$ with respect to the scalar product
\begin{align}
 \langle \xi_n(r),\xi_m(r)\rangle=\int_0^{r_\mathrm{nuc}} r^2 \xi_n(r)\xi_m(r) \mathrm{d}r=\delta_{n m}.
\end{align}
To model a driving force acting on the system we can expand the Dirac delta function on the surface
\begin{align}\label{deltaExpand}
 \delta(r-r_\mathrm{nuc})=-(2\,r_\mathrm{nuc})^\frac{1}{2}\sum_{n=1}^\infty (-1)^n \xi_n(r)
\end{align}
and the identity function
\begin{align}\label{idExpand}
 \mathrm{id}(r)=-\frac{3 (2\,r_\mathrm{nuc}^5)^\frac{1}{2}}{\pi^2}\sum_{n=1}^\infty \frac{(-1)^n}{n^2} \xi_n(r)
\end{align}
in terms of this basis. The spatial delta function at the surface can be used in combination with a temporal delta function
to model an instantaneous impulse applied to the surface, or in combination with a harmonic temporal function to drive the surface
periodically. The identity function in combination with a harmonic temporal function models the breathing mode of a homogenously
charged sphere. To first order, Gauss' law predicts an electric force proportional to the radius resulting from any contraction
or expansion of the sphere.

\section{Solving the Equation of Motion}

According to (\ref{diffEqn}) and (\ref{omegaN}) the basis functions fulfil the equation
\begin{align}
\omega_n^2\xi_n(r)-\boxtimes\xi_n(r)=0
\end{align}
where $\boxtimes$ is a differential operator in $r$.
Making the ansatz
\begin{align}
 \xi(t,r)=\sum_{n=1}^\infty T_n(t)\xi_n(r)
\end{align}
and plugging it into the equation of motion subject to an external force
\begin{align}
 \ddot\xi(t,r)-\boxtimes\xi_n(t,r)=f(t,r)
\end{align}
we can expand in the basis $(\xi_n)$ to obtain the independent equations
\begin{align}\label{eqnsOM}
 \ddot T_n(t)-\omega_n^2 T_n(t)=f_n(t),
\end{align}
where
\begin{align}
 f(t,r)=\sum_{n=1}^\infty f_n(t) \xi_n(r).
\end{align}
If the force is applied just for an instant at $t=0$ we can write $f_n(t)=f_n \delta(t)$,
and assuming the system initially unperturbed ($T_n(t)=0$ for $t<0$) the solution to (\ref{eqnsOM}) is
\begin{align}
 T_n(t)=\frac{f_n}{\omega_n}\theta(t) \sin(\omega_n t).
\end{align}
If the force is applied at a frequency $\Omega$ starting at $t=0$ we can write $f_n(t)=f_n \theta(t) \sin(\Omega\,t)$,
and assuming the system initially unperturbed the solution to (\ref{eqnsOM}) is
\begin{align}
 T_n(t)=f_n \theta(t)\frac{\omega_n \sin(\Omega\,t)-\Omega\,\sin{\omega_n t}}{\omega_n^3-\omega_n\Omega^2}.
\end{align}
Using for $f_n$ now the coefficients of (\ref{deltaExpand}) or (\ref{idExpand}), we can construct various dynamical
situtations in the compressed atom.

\section{Conclusions}

Using the formalism developed above, it is now possible to investigate the response of the electron gas to the nuclear breathing
mode for different system sizes from atoms and giant atoms to neutron stars. The proton density should be determined from beta-equilibrium,
while the frequency of the breathing mode has to be inferred from studies of nuclear incompressibility such as \cite{blaizot,wang}. An
investigation has to take into account the energy budget of the intitial breathing mode, comparing it to the energy transferred to electron
gas and electric field, to determine if dampening of the former occurs first.

\section*{Acknowledgements}
We thank the National Academy of Sciences of Belarus for their hospitality. H. Ludwig is supported by the Erasmus Mundus Joint Doctorate Program by Grant
Number 2012-1710 from the EACEA of the European Commission.
\\
\\
\\
H. Ludwig: hendrik.ludwig@icranet.org\\
R. Ruffini: ruffini@icra.it\\
S.-S. Xue: xue@icra.it\\
all authors: P.zza della Republica 10, I-65122 Pescara, Italy


\newpage
\begin{thebibliography}{}
\bibitem{sauter}
F.~Sauter, Z.~Phys. {\bf 82}, 742 (1931).

\bibitem{ludwig}
H.~Ludwig and R.~Ruffini and S.-S.~Xue, {\it arXiv:1402.3468} (2014).

\bibitem{wang}
Wang, C.S.~ and K.C.~Chung and A.J.~Santiago, Phys.~Rev.~C {\bf 55}, 2844 (1997).

\bibitem{preparata}
G.~Preparata and R.~Ruffini and S.-S.~Xue, Astron.~Astrophys. {\bf 338}, L87 (1998).

\bibitem{ipp}
A.~Ipp and J.~Evers and C.H.~Keitel and K.Z.~Hatsagortsyan, Phys.~Lett.~B {\bf 702}, 383 (2011).

\bibitem{blaizot}
J.P.~Blaizot and J.F.~Berger and J.~Decharg{\'e} and M.~Girod, Nuc.~Phys.~A {\bf 591}, 435 (1995).
\end{thebibliography}

\newpage

Fig.1. The basis functions $\xi_n$ for $n=1,2,3,4$ are shown for $r_\mathrm{nuc}=10$, corresponding to $Z=1000$ and $\Delta=1$. For ascending $n$,
the lines are solid, dashed, dotted and dash-dotted, respectively.\\

\end{document}